\documentclass[article, prl, nobibnotes, secnumarabic, amssymb, superscriptaddress, twocolumn, aps,longbibliography]{revtex4-2}

\usepackage{graphicx}
\usepackage{amsmath}
\usepackage{upgreek}
\usepackage{color}
\usepackage{setspace}
\usepackage{hyperref}
\hypersetup{colorlinks,breaklinks,
            urlcolor=[rgb]{0,0,0.64},
            linkcolor=[rgb]{0,0,0.64},
            citecolor=[rgb]{0,0,0.64}}
\usepackage{nccmath}
\usepackage{amssymb}
\usepackage{lineno}

\newcommand{\MIT}{Massachusetts Institute of Technology, Department of Physics, Cambridge, MA 02139, USA.}
\newcommand{\MITC}{Massachusetts Institute of Technology, Department of Electrical Engineering and Computer Science, Cambridge, MA 02139, USA.}
\newcommand{\TDLI}{Tsung-Dao Lee Institute, Shanghai Jiao Tong University, Shanghai 200240, China.}
\newcommand{\Cal}{University of California at Berkeley, Department of Chemistry, Berkeley, CA 94720, USA.}
\newcommand{\ICQM}{International Center for Quantum Materials, School of Physics, Peking University, Beijing 100871, China.}

\newcommand{\StanfordAP}{Department of Applied Physics, Stanford University, Stanford, CA 94305, USA.}
\newcommand{\SSRF}{Shanghai Synchrotron Radiation Facility, Shanghai Advanced Research Institute, Chinese Academy of Sciences, Shanghai 201204, China.}

\newcommand{\SLAC}{SLAC National Accelerator Laboratory, Menlo Park, CA 94025, USA.}
\newcommand{\Emory}{Department of Chemistry, Emory University, Atlanta, GA 30322, USA.}
\newcommand{\IOP}{Beijing National Laboratory for Condensed Matter Physics and Institute of Physics, Chinese Academy of Sciences, Beijing 100190, China.}
\newcommand{\BAQIS}{Beijing Academy of Quantum Information Sciences, Beijing 100913, China.}
\newcommand{\HKU}{Department of Physics, Hong Kong University of Science and Technology, Clear Water Bay, Hong Kong, China.}
\newcommand{\SPA}{School of Physics and Astronomy, Shanghai Jiao Tong University, Shanghai 200240, China.}
\newcommand{\ZIAS}{Zhangjiang Institute for Advanced Study, Shanghai Jiao Tong University, Shanghai 200240, China.}

\newcommand{\papertitle}{Coexistence of interacting charge density waves in a layered semiconductor} 

\newcommand{\ConI}{{\fontfamily{cmtt}\selectfont Con1}}
\newcommand{\ConII}{{\fontfamily{cmtt}\selectfont Con2}}
\newcommand{\ConIII}{{\fontfamily{cmtt}\selectfont Con3}}

\begin{document}

\title{\papertitle}

\author{B.~Q.~Lv}
\thanks{These authors contributed equally to this work: B.~Q.~Lv and Alfred~Zong.}
\affiliation{\TDLI}
\affiliation{\MIT}
\affiliation{\SPA}
\affiliation{\ZIAS}
\author{Alfred~Zong}
\thanks{These authors contributed equally to this work: B.~Q.~Lv and Alfred~Zong.}
\affiliation{\MIT}
\affiliation{\Cal}
\author{Dong~Wu}
\affiliation{\BAQIS}
\author{Zhengwei~Nie}
\affiliation{\IOP}
\author{Yifan~Su}
\affiliation{\MIT}
\author{Dongsung~Choi}
\affiliation{\MITC}
\author{Batyr~Ilyas}
\affiliation{\MIT}
\author{Bryan~T.~Fichera}
\affiliation{\MIT}
\author{Jiarui~Li}
\affiliation{\SLAC}
\affiliation{\StanfordAP}
\author{Edoardo~Baldini}
\affiliation{\MIT}
\author{Masataka~Mogi}
\affiliation{\MIT}
\author{Y.-B.~Huang}
\affiliation{\SSRF}
\author{Hoi~Chun~Po}
\affiliation{\HKU}
\author{Sheng~Meng}
\affiliation{\IOP}
\author{Yao~Wang}
\affiliation{\Emory}
\author{N.~L.~Wang}
\affiliation{\BAQIS}
\affiliation{\ICQM}
\author{Nuh~Gedik}
\email[Correspondence to: ]{gedik@mit.edu}
\affiliation{\MIT}

\date{\today}

\begin{abstract}
Coexisting orders are key features of strongly correlated materials and underlie many intriguing phenomena from unconventional superconductivity to topological orders. 
Here, we report the coexistence of two interacting charge-density-wave (CDW) orders in EuTe$_4$, a layered crystal that has drawn considerable attention owing to its anomalous thermal hysteresis and a semiconducting CDW state despite the absence of perfect FS nesting. By accessing unoccupied conduction bands with time- and angle-resolved photoemission measurements, we find that mono- and bi-layers of Te in the unit cell host different CDWs that are associated with distinct energy gaps. The two gaps display dichotomous evolutions following photoexcitation, where the larger bilayer CDW gap exhibits less renormalization and faster recovery. Surprisingly, the CDW in the Te monolayer displays an additional momentum-dependent gap renormalization that cannot be captured by density-functional theory calculations. This phenomenon is attributed to interlayer interactions between the two CDW orders, which account for the semiconducting nature of the equilibrium state. Our findings not only offer microscopic insights into the correlated ground state of EuTe$_4$ but also provide a general non-equilibrium approach to understand coexisting, layer-dependent orders in a complex system.
\end{abstract}

\maketitle

\begin{figure}[htb!]
    \includegraphics[width=\columnwidth]{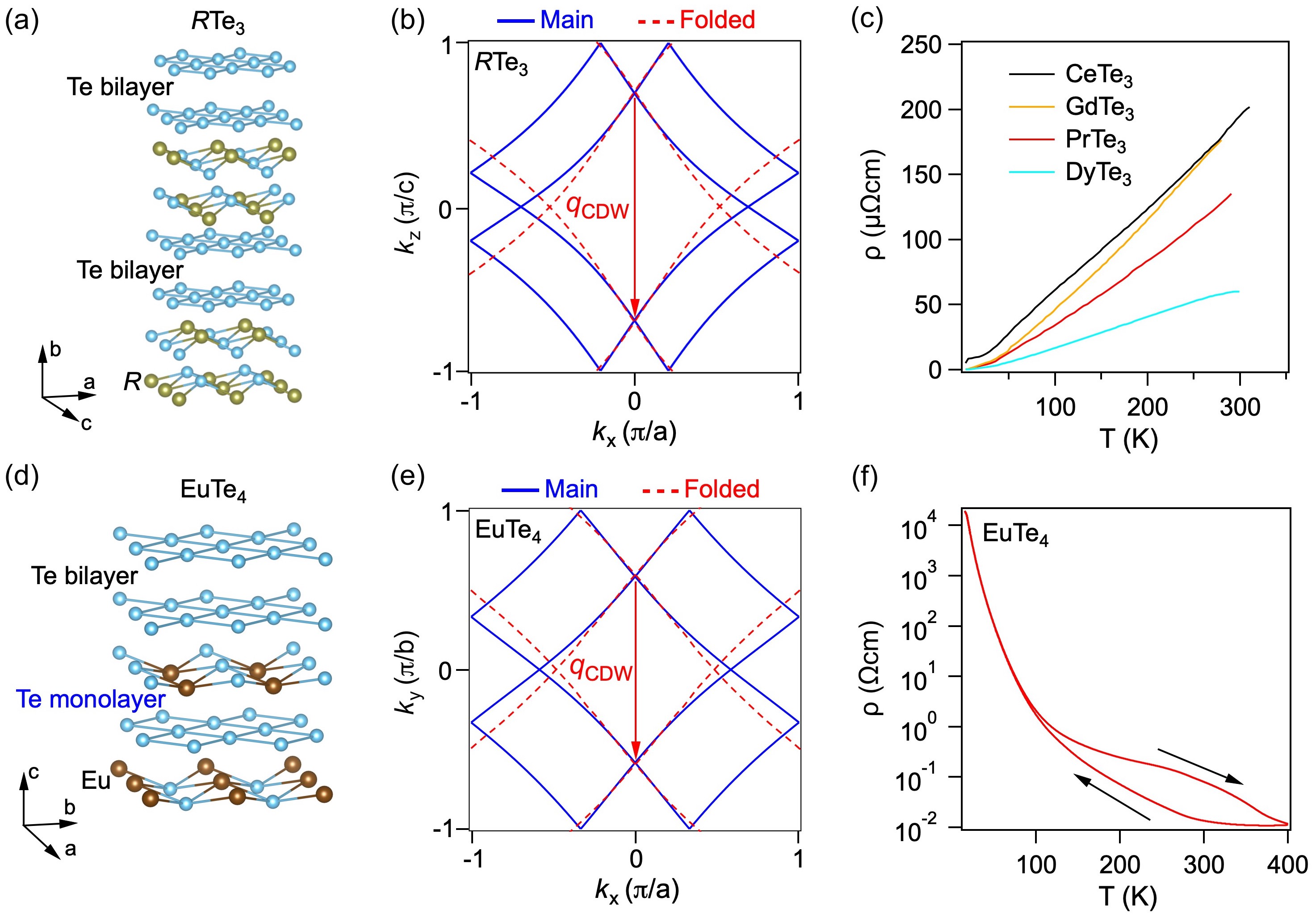}
    \caption{Comparison between charge density waves in \textit{R}Te$_\text{3}$ and EuTe$_\text{4}$. 
    (a)~Crystal structure of $R$Te$_3$, featuring Te bilayers whose states are closest to the Fermi level. 
    (b)~Calculated normal-state FS (blue lines) of $R$Te$_3$ based on the tight-binding model. The blue solid and red dashed lines represent the original and folded bands, respectively. Note that only parts of folded bands are plotted for clarity.  (c)~Temperature-dependent electrical resistivity of $R$Te$_3$, showing metallic behavior in the CDW state. The CDW transition temperatures of all the compounds shown are above 300~K. (d)~Crystal structure of EuTe$_4$. 
    (e)~Tight-binding normal-state FS of EuTe$_4$, following the same convention as in panel (b). (f)~Temperature-dependent electrical resistivity of EuTe$_4$ in the CDW state, showing semiconducting behavior and a giant thermal hysteresis. Panels~(c) and (f) are extracted from ref.~\cite{Iyeiri2003} and ref.~\cite{LvPRL2021}, respectively.}
\label{fig1}
\end{figure}

Quantum materials driven by nonperturbative correlations display rich phase diagrams thanks to multiple instabilities at similar energy and time scales \cite{Keimer2015}. The microscopic understanding of macroscopic manifestations of the resulting coexistent orders has been a central subject in condensed matter physics. In solids, order formation due to symmetry-breaking is often accompanied by the appearance of an energy gap near the Fermi level, where the gap size is proportional to the amplitude of the order parameter. A prominent example is the charge-density-wave (CDW) order. 
Historically, Fermi surface (FS) nesting --- the matching of sections of the FS to some other parts by a single wave vector $\mathbf{q}$ --- was suggested as the main driver for CDW formation in one-dimensional chains \cite{Peierls1955,Gruner1994}, leading to a phase transition from the metallic to semiconducting state. 
For quasi-two-dimensional materials, however, a perfect nesting condition can seldom be fulfilled due to the complex FS topology, and other mechanisms are often at play for the CDW formation \cite{Johannes2008,Zhu2015}. Hence, in quasi-2D materials, the CDW energy gap usually does not open everywhere on the FS and the system remains metallic even in the CDW state. One well-established example is the family of $R$Te$_3$ ($R = $ lanthanide except Pm, Eu, Yb, and Lu) \cite{Dimasi1995,Malliakas2005,RuThesis,Yumigeta2021}, which is known for the CDW states originating from the conducting Te bilayers [Fig.~\ref{fig1}(a)]. The normal-state FS can be well approximated by a tight-binding model that considers the Te bilayers \cite{Brouet2008}, as shown in Fig.~\ref{fig1}(b). One can see that the FS favors a good but not perfect nesting, leading to the metallic CDW state with a partially gapped FS, as evidenced by metallic behavior in electrical transport [Fig.~\ref{fig1}(c)] and remnant FS seen in angle-resolved photoemission spectroscopy (ARPES) \cite{Gweon1998,Iyeiri2003,Brouet2008}.

The recently synthesized Eu-based telluride EuTe$_4$, where Eu is divalent instead of the trivalent lanthanide in $R$Te$_3$, has stimulated intense interest due to its unique electrical transport property: an anomalously large thermal hysteresis embedded in a semiconducting CDW state \cite{Wu2019,LvPRL2021,Pathak2022,ZhangQQ2023,Zhangchen2022,Rathore2023,Liu2023}, shown in Fig.~\ref{fig1}(f). EuTe$_4$ shares key structural motifs with $R$Te$_3$, both consisting of an alternate stacking of the insulating $R$Te spacers and the nearly-square Te nets [Fig.~\ref{fig1}(a),(d)]. 
As expected, the calculated normal-state FS of EuTe$_4$ exhibits imperfect nesting, as illustrated in Fig.~\ref{fig1}(e). These similarities point to a metallic CDW state, which is also suggested by density-functional theory (DFT) calculations \cite{Wu2019,Pathak2022}. However, previous ARPES studies revealed a fully gapped FS \cite{LvPRL2021,ZhangQQ2023,Zhangchen2022}. The stark contrast raises the puzzle of why the CDW state of EuTe$_4$ is semiconducting. Structurally, the apparent difference is that EuTe$_4$ contains both Te monolayers and bilayers in a single unit cell, hinting at the possibility of coexistence of layer-dependent CDW orders. Therefore, one important clue is the additional interactions between the two CDW orders. Yet, despite deploying multiple experimental probes \cite{LvPRL2021,ZhangQQ2023}, it remains unclear whether more than one CDW order exists in EuTe$_4$ in the first place.
In this work, we leverage the capability of time-resolved ARPES (tr-ARPES)  \cite{Gedik2017,Sobota2021,RMPultrafast2021,Boschini2023} to distinguish and probe the optical excitation of two CDW orders in EuTe$_4$. Our findings suggest that the interlayer CDW coupling underpins the semiconducting nature of EuTe$_4$, and our methodology offers a promising route for visualization and ultrafast manipulation of coexisting orders in other quasi-2D systems.

\begin{figure}[tb!]
    \includegraphics[width=\columnwidth]{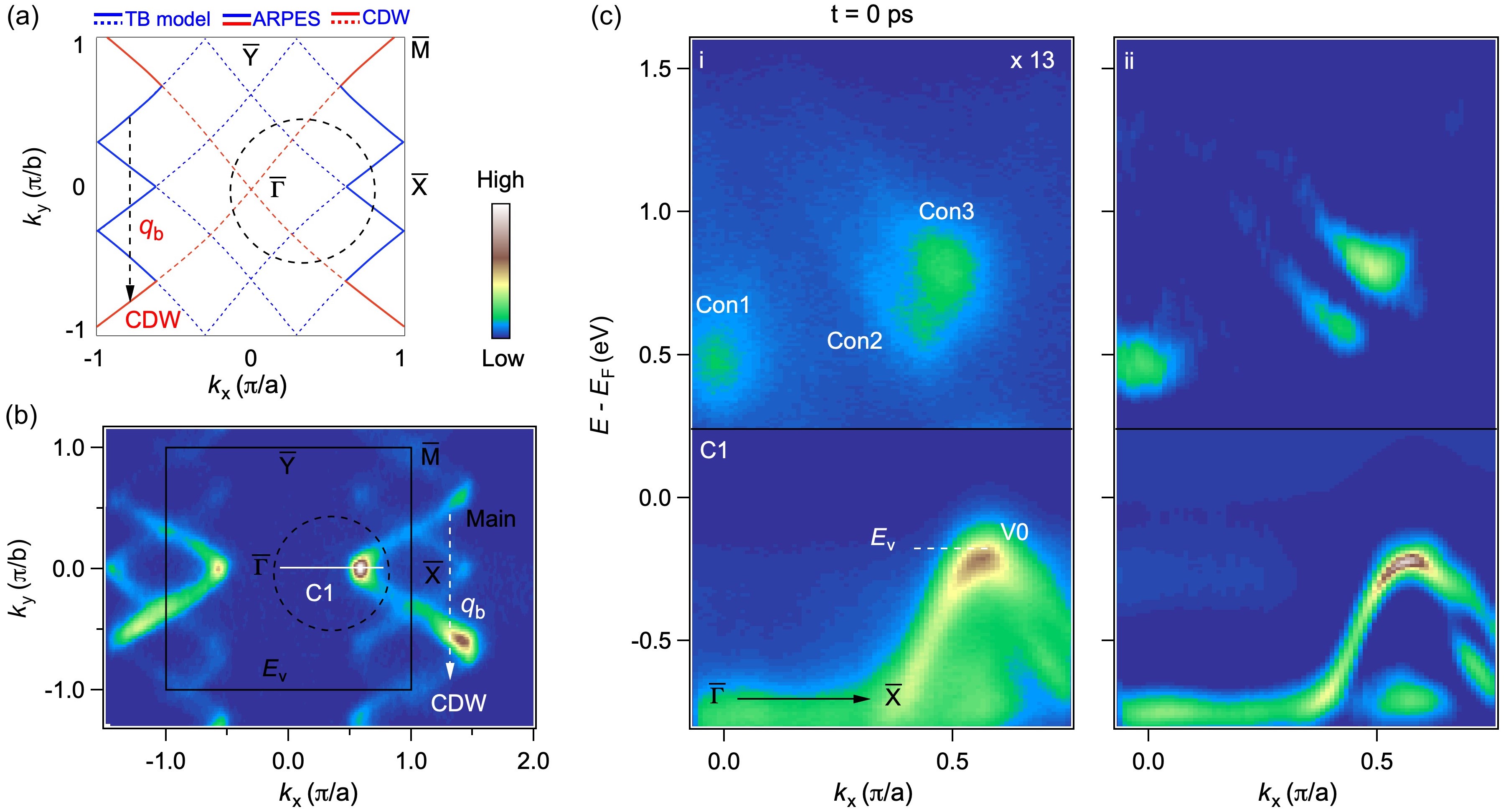}
    \caption{Electronic structure and intrinsic CDW gaps of EuTe$_\text{4}$. (a)~Tight-binding original (blue lines) and folded (red lines) Fermi surface. The solid lines represent the observed bands at $E_v$ [shown in panel~(b)]. The black dashed circle indicates the detection area of our tr-ARPES measurements. (b)~ARPES constant energy map at $E_v$, measured with 90~eV, right circularly polarized light. (c)~ARPES (i) and curvature (ii) intensity plots along the C1 direction [labeled in panel(b)], measured at $t= 0$~ps. 
    The probe photon energy was 10.75~eV, and polarization was linear horizontal (LH, perpendicular to the photoemission plane). The pump photon energy was 1.55~eV, the polarization was LH, and the fluence was 0.16~mJ/cm$^2$.}
\label{fig2}
\end{figure}

We first examine the occupied valence bands of EuTe$_4$ in its CDW state. The calculated normal-state FS, as shown by the blue curves in Fig.~\ref{fig1}(e), is mainly composed of two perpendicular sets of quasi-parallel curves. In the CDW state, these curves are folded along the $k_y$ direction, leading to additional Fermi pockets in the Brillouin zone [see red curves in Fig.~\ref{fig2}(a)]. We measured the fermiology in the CDW state by high-resolution synchrotron-based ARPES at 20~K. Figure~\ref{fig2}(b) displays a representative constant-energy map at $E_v$ [indicated in Fig.~\ref{fig2}(c)], where both the original bands and the folded CDW bands are resolved. The measurement in Fig.~\ref{fig2}(b) is largely consistent with the tight-binding prediction in Fig.~\ref{fig2}(a) except for the missing pockets at $|k_x| \lesssim 0.5\pi/a$ [indicated by dashed curves in Fig.~\ref{fig2}(a)], which are the consequence of large energy gaps in the CDW state, as we will quantify later.

\begin{figure*}[tb!]
    \includegraphics[width=1.0\textwidth]{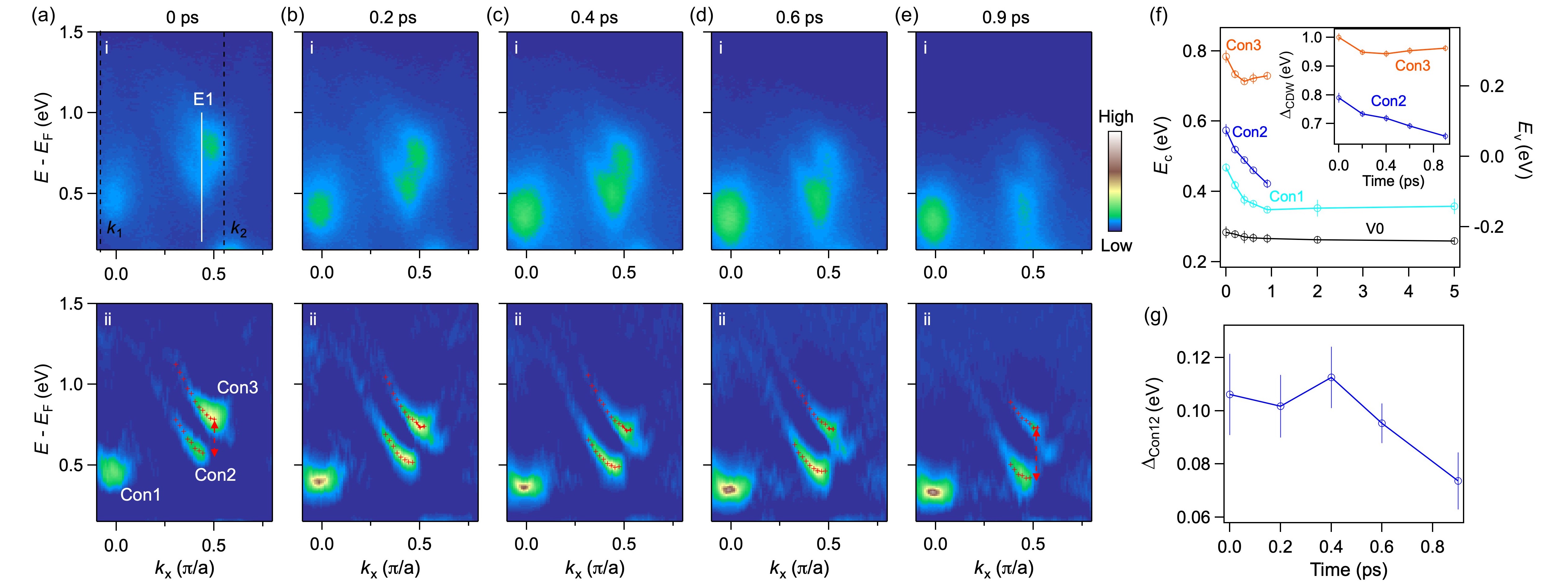}
    \caption{Temporal evolution of conduction bands and CDW gaps. (a)--(e)~ARPES (i) and curvature (ii) intensity plots along the C1 direction at $t= 0$, 0.2, 0.4, 0.6, and 0.9~ps, respectively. The red symbols are peak positions of \ConII{} and \ConIII{} bands, obtained from the Lorentzian fitting of the energy distribution curves at each $k_x$point and at each delay time. 
    (f)~Extracted time-dependent conduction band bottoms. The inset shows the corresponding CDW gaps determined from the difference between the bottom of \ConII{} or \ConIII{} and the top of V0. (g)~Time-dependent band-bottom difference of \ConI{} and \ConII{}. The error bars represent the statistical uncertainty derived from the fitting procedure.}
\label{fig3}
\end{figure*}

To discern the possible existence of two CDWs, information about unoccupied single-particle states must be characterized, which gives a more quantitative measure of the CDW energy gap and hence reflects the intrinsic order parameter amplitude. These unoccupied states are difficult to detect in equilibrium ARPES due to the large gap size compared to thermal energy at experimentally accessible temperatures up to $\sim400$~K \cite{LvPRL2021}; inverse ARPES would also be unsuitable due to the coarse energy resolution. The limitations of the equilibrium techniques necessitate the use of tr-ARPES to differentiate CDW orders in EuTe$_4$. With this in mind, we conducted tr-ARPES measurements using an above-gap infrared (1.55~eV) pump laser and an extreme-ultraviolet (XUV, 10.75~eV) probe laser (see \cite{Remark1} for experimental details). 
Figure~\ref{fig2}(c) shows the measured transient electronic structure along $\Gamma$--X under an incident fluence of 0.16~mJ/cm$^2$ at $t \sim 0$~ps (i.e., at pump-probe temporal overlap). In this weak perturbation regime such that Floquet states can be neglected, pump pulses are expected to minimally disturb the CDW structure at $t \sim 0$~ps. As a result, the intensity that indicates populated states above $E_F$ more closely resembles the unoccupied states in equilibrium compared to spectra at later time delays \cite{Schmitt_2011}, giving us access to the unoccupied band structure that is otherwise inaccessible in equilibrium. 

Three spectral features, labeled as \ConI{}, \ConII{}, and \ConIII{} in Fig.~\ref{fig2}(c), can be identified above $E_F$, indicating three conduction-band bottoms in the CDW state. \ConI{} resides at the $\Gamma$ point while \ConII{} and \ConIII{} are close to each other in crystal momentum along the $\Gamma$--X direction but are energetically distinct. To trace the origins of these unoccupied bands, we compare Fig.~\ref{fig2}(c) with Fig.~\ref{fig2}(a) and the calculated normal-state electronic structure in Fig.~S6(a). It is evident that \ConI{} can only be assigned to the folded CDW band since no normal-state bands lie within an energy window of $\pm1$~eV at $\Gamma$. By contrast, \ConII{} and \ConIII{} are parts of the normal-state main bands with a gap opening near $E_F$. By calculating the difference between the energy minima of \ConII{} and \ConIII{} and the energy maximum of the valence band top [labeled as V0 in Fig.~\ref{fig2}(c)], we estimate the corresponding CDW gaps along $\Gamma$--X to be 0.8~eV and 1.0~eV, respectively. The presence of two distinct single-particle gaps hence demonstrates the coexistence of two CDW orders.

Spectroscopically, separated bands like \ConII{} and \ConIII{} do not always guarantee two distinct orders because scenarios such as spin-orbit coupling can also lead to band splitting. To rule out other scenarios and to pin down the coexistence of the two CDW orders, we study the evolution of their spectral functions after photoexcitation. Figures~\ref{fig3}(a)--(e) show the snapshots of photoemission spectra above $E_F$ at five representative pump-probe delay points (first row) and their respective curvature plots that highlight the band positions (second row). Besides the time-dependent spectral weights due to photo-induced carrier excitation and recombination, the energy positions of all three conduction bands \ConI{}--\ConIII{} shift down immediately after the pump pulse arrival, indicating an overall non-equilibrium suppression of the CDW order. Importantly, there is a larger energy separation between \ConII{} and \ConIII{} in Fig.~\ref{fig3}(e) compared to Fig.~\ref{fig3}(a) (highlighted by the dashed arrows), suggesting two different evolution pathways for the two bands. To quantify this distinction, we fit the energy distribution curves (EDCs) between $k_1$ and $k_2$ labeled in Fig.~\ref{fig3}(a)i with Lorentzian functions (see \cite{Remark1}). The extracted momentum-dependent EDC peak positions of \ConII{} and \ConIII{} are superimposed on the curvature plots [Fig.~\ref{fig3}(a)ii--\ref{fig3}(e)ii] with red markers, which very well match the band intensity and hence affirm the validity of both the EDC fits and the curvature plots. We summarize the fitting results in Fig.~\ref{fig3}(f). \ConII{} and \ConIII{} clearly exhibit different dynamics. The bottom of \ConIII{} decreases from 0.78~eV at $t = 0$ to 0.71~eV at 0.4~ps, and then starts to recover after 0.4~ps. By contrast, the bottom of \ConII{} keeps decreasing until 0.9~ps. After 0.9~ps, both \ConII{} and \ConIII{} are undetectable [see Fig.~S4] due to fast excited carrier relaxation and hence the disappearance of spectral weights. These distinct dynamics lend further proof that \ConII{} and \ConIII{} come from two different orders.

We can associate the above observations of two CDW gap sizes and two qualitatively different photoinduced dynamics with the unique CDW instabilities and structural motifs in EuTe$_4$, which feature both Te monolayers and Te bilayers. Due to the quasi-2D nature of EuTe$_4$, strong hybridization of states between the mono- and bi-layers are limited, so it is reasonable to attribute the two CDW orders to the monolayer and bilayer Te sheets. To obtain the specific assignment of \ConII{} and \ConIII{}, we note that within a Te bilayer, the additional intra-bilayer interaction can further enhance the broken symmetry order, as has been verified by our DFT results in Fig.~S6. Hence, the CDW gap associated with Te bilayers is expected to be larger, and we assign \ConIII{} and \ConII{} to the bilayer and monolayer CDW, respectively. This assignment was further supported by the observed dichotomous temporal evolutions in Fig.~\ref{fig3}. 
The stronger bilayer CDW experiences less suppression and hence a faster recovery (\ConIII{}), whereas the weaker monolayer CDW has a larger gap suppression and delayed recovery [Fig.~\ref{fig3}(f)]. These characteristics not only provide strong evidence for the existence of monolayer and bilayer CDWs but also suggest that photoexcitation can be an effective means for charting separate pathways of their dynamics due to the different CDW strengths.

\begin{figure}[htb!]
    \includegraphics[width=1\columnwidth]{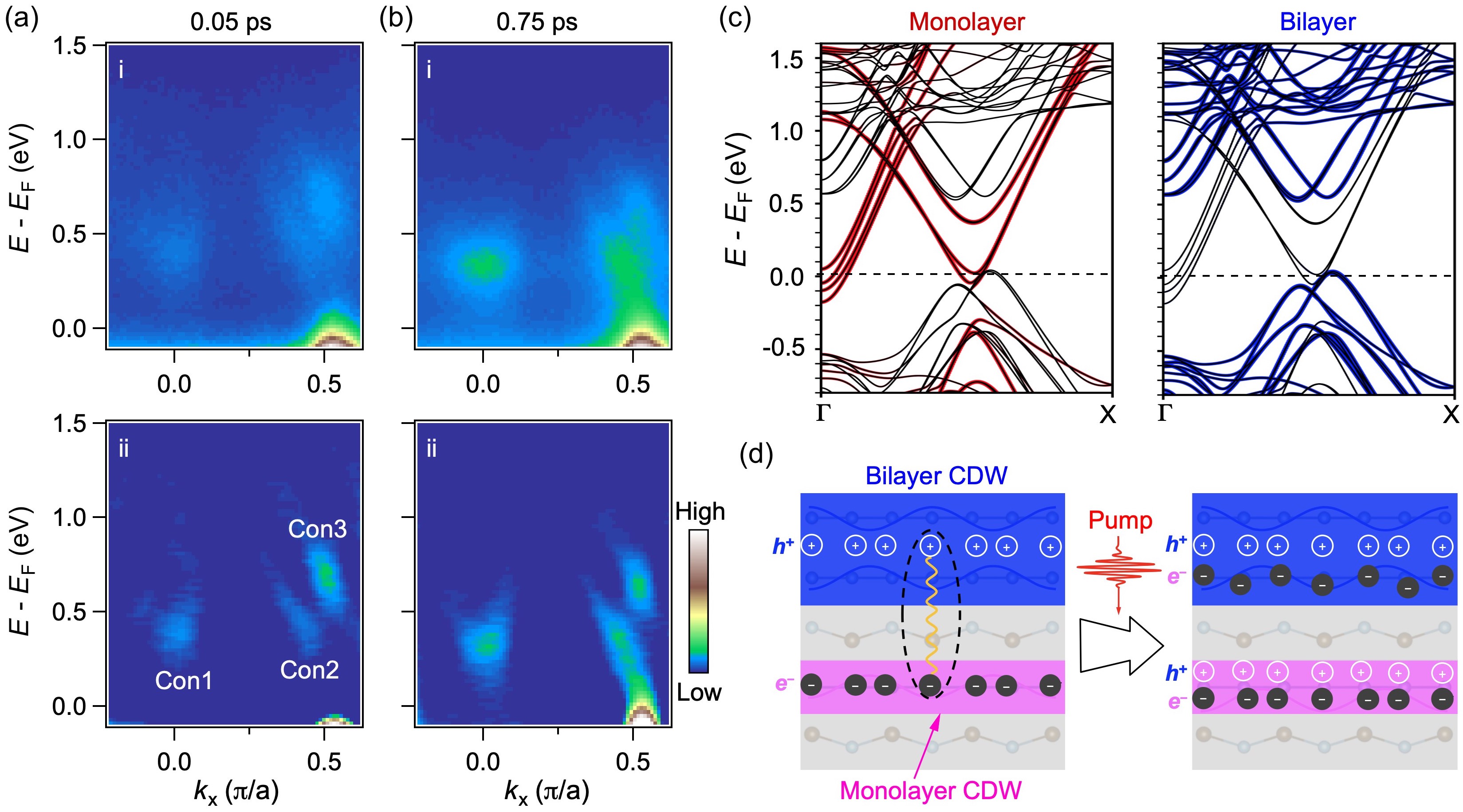}
    \caption{The interplay of intertwined CDW states.
    (a),(b)~ARPES (i) and curvature (ii) intensity plots along the $k_x$ direction at $t= 0.05$ and 0.75~ps, respectively. The pump 
    fluence was 1.2~mJ/cm$^2$. (c)~The calculated band structure along the $\Gamma$--X direction based on the ($1\times3\times2$) superlattice and the GGA type of the exchange-correlation potential. The red and blue colors represent the weight of $p_x$ and $p_y$ orbitals from the Te monolayer and bilayer, respectively. (d)~Schematic of the electron-hole interactions between CDWs residing in the monolayer and bilayer Te sheets. The white circles and black dots represent the holes and electrons, respectively. The orange curves denote the interactions, and the black-dotted ellipse represents an electron-hole pair.}
\label{fig4}
\end{figure}

We next discuss the behavior of \ConI{}, which, unlike \ConII{} and \ConIII{}, is a part of the folded bands that are only present in the CDW state. We assign \ConI{} to the CDW residing in the monolayer Te sheet for two reasons. First, \ConI{} is a global conduction band minimum as directly measured in Fig.~\ref{fig3}(a) and as evidenced by the long population lifetime in Fig.~S4, which results from the slow equilibration of excited electrons located at the conduction band bottom of a semiconductor \cite{PRLKanasaki2014}. Second, \ConI{} displays a similar dynamical evolution as \ConII, namely, both keep shifting downward toward larger binding energy over $\sim 1$~ps. The main difference between \ConI{} and \ConII{} is the magnitude of the shift, where \ConI{} exhibits a smaller downshift on average compared to \ConII{}. As a result, the difference between the energy bottoms of \ConI{} and \ConII{} decreases as a function of time, as shown in Fig.~\ref{fig3}(g). Such discrepancy is more prominent under stronger laser excitation. In Fig.~\ref{fig4}(a) and (b), we present the transient electronic structure at $t = 0.05$ and 0.75~ps after excitation with 1.2~mJ/cm$^2$, 800-nm pump pulses. Under such a strong perturbation, the energy gap between \ConII{} and V0 almost completely collapses due to the large downshift of \ConII{}. By contrast, \ConI{} shows a much smaller downshift and it is clear that the energy gap does not vanish at the $\Gamma$ point.

The observation of disproportionate gap renormalization of \ConI{} and \ConII{} is unexpected because it conflicts with the previous understanding of EuTe$_4$ that there is only one order parameter responsible for gap opening. In other words, one should expect a proportional downshift of \ConI{} and \ConII{} if the monolayer CDW has a single CDW order parameter. To understand this behavior,  we performed detailed DFT calculations based on the $1 \times 3 \times 2$ superlattice, which is a good approximation of the in-plane incommensurate and out-of-plane commensurate CDW structure \cite{LvPRL2021,Wu2019}. The DFT results, in good agreement with previous reports \cite{Wu2019,PRB2022}, are summarized in Fig.~\ref{fig4}(c) and \cite{Remark1}. One can see that the Te monolayer indeed has a relatively weaker CDW, resulting in several conduction band bottoms, supporting our assignment of \ConI{}, \ConII{}, and \ConIII{} in Fig.~\ref{fig3}. On the other hand, besides an overall shift of the chemical potential, the major discrepancy between the DFT results and the measured bands in Fig.~\ref{fig2}(c) is that DFT yields a metallic FS, suggesting that additional electron correlations beyond those captured by DFT must be considered to explain the observed semiconducting CDW, especially for the monolayer CDW gap at \ConII{}.

Motivated by the recent investigations of the insulating state in 1$T$-TaS$_\text{2}$ induced by interlayer coupling \cite{wenchenh2021,Butler2020,Wang2020,Ritschel2018,LeeS2019}, we speculate that the experimentally observed gap in EuTe$_4$ arises from mono- and bi-layer CDW couplings. 
A direct observation from the DFT-calculated band structure is that the low-energy conduction and valence bands are primarily contributed by monolayer and bilayer Te orbitals, respectively (Fig.~S6). The energy proximity and spatial separation of electrons and holes make the material susceptible to charge transfer between these nominally charge-neutral Te layers\,\cite{halperin1968possible}. Since DFT is a ground-state theory and the functionals do not include excitonic effects, we can compensate for this missing interlayer interaction by an additional term in the Hamiltonian, $Vn^{(\alpha)} n^{(\beta)}$, where $n^{(\alpha)}$ and $n^{(\beta)}$ correspond to the electron density operators at the monolayer and bilayer Te bands, and $V$ represents the attractive interband coupling. 
A direct consequence is the separation of these types of CDW bands [red and blue curves in Fig.~\ref{fig4}(c)] caused by the Hartree part of this interaction. Furthermore, this interaction favors the formation of inter-band excitons, whose condensation opens an excitonic gap. Both gapping mechanisms are directly driven by, and therefore proportional to, $V$, without requiring a structural transition. It is predicted that this effective interaction can be screened by carriers in both layers, leading to a reduction of the aforementioned gap. This is consistent with the experimentally observed shift of \ConII{} [see Fig.~\ref{fig4}(d)] with the presence of light-induced photocarriers. At the same time, the \textit{intra}layer CDW orders remain unchanged, manifesting as the persistence of the \ConI{}, highlighting the distinct origins of the CDW gap and the interband interaction-driven gap.

Our time-domain spectroscopic investigations of the CDW state in EuTe$_4$ unveil the following key results. (i)~In a single bulk crystal, the Te monolayer and bilayer host distinct orders, with the bilayer experiencing a stronger CDW distortion due to the additional intra-bilayer interactions. (ii)~As a result, the bilayer CDW is less renormalized and recovers faster under small perturbations by light. (iii)~The monolayer \ConI{} and \ConII{} conduction bands exhibit disproportionate temporal evolutions, deviating from the typical dynamics of a single-order parameter. We interpret this deviation and the related semiconducting nature of the CDW as a result of additional mono-bilayer interactions. These findings provide crucial information on the dynamics and interplay of coexisting CDW orders, highlighting the importance of interlayer coupling in semiconducting quasi-2D CDW systems. As a newly discovered CDW material, EuTe$_4$ also offers a rich platform for understanding and manipulating the layer-specific CDW orders with a variety of external parameters such as temperature and ultrashort light pulses, which can result in anomalous electrical transport and persistent hidden states \cite{LvPRL2021,Liu2023}. The above knowledge of coexisting CDW orders can also be generalized to other layered systems, such as multilayer cuprates \cite{Damascelli2003}, and our study thereby offers another route for elucidating the novel physics of coexisting orders in these quantum many-body systems.

\begin{acknowledgments}
We thank B.~V.~Fine, A.~V.~Rozhkov, J.-Z.~Zhao, and D.~Azoury for fruitful discussions. We acknowledge support from the U.S. Department of Energy, Office of Science, Office of Basic Energy Sciences, DMSE (instrumentation and data taking), the National Science Foundation under Grant No.~NSF DMR-1809815 (data analysis), and the Gordon and Betty Moore Foundation's EPiQS Initiative grant GBMF9459 (manuscript writing). B.Q.L. acknowledges from the TDLI starting up grant, the National Natural Science Foundation of China (23Z990202580), the Ministry of Science and Technology of China (2023YFA1407400), the Shanghai Natural Science Fund for Original Exploration Program (23ZR1479900), and the Shanghai Talent Program.  A.Z. acknowledges support from the Miller Institute for Basic Research in Science.
D.W. and N.L.W. acknowledge support from the National Natural Science Foundation of China (No.~11888101), and the National Key Research and Development Program of China (No.~2022YFA1403901). 
Y.B.H. acknowledges support from the National Key Research and Development Program of China (2017YFA0403401) and the National Natural Science Foundation of China (U1875192, U1832202). Y.W. acknowledges support from U.S. Department of Energy, Office of Science, Basic Energy Sciences, under Early Career Award No. DE-SC0024524. E.B. acknowledges additional support from the  Swiss  NSF  under fellowships  P2ELP2-172290  and  P400P2-183842.
\end{acknowledgments}


\begin{thebibliography}{34}%
\makeatletter
\providecommand \@ifxundefined [1]{%
 \@ifx{#1\undefined}
}%
\providecommand \@ifnum [1]{%
 \ifnum #1\expandafter \@firstoftwo
 \else \expandafter \@secondoftwo
 \fi
}%
\providecommand \@ifx [1]{%
 \ifx #1\expandafter \@firstoftwo
 \else \expandafter \@secondoftwo
 \fi
}%
\providecommand \natexlab [1]{#1}%
\providecommand \enquote  [1]{``#1''}%
\providecommand \bibnamefont  [1]{#1}%
\providecommand \bibfnamefont [1]{#1}%
\providecommand \citenamefont [1]{#1}%
\providecommand \href@noop [0]{\@secondoftwo}%
\providecommand \href [0]{\begingroup \@sanitize@url \@href}%
\providecommand \@href[1]{\@@startlink{#1}\@@href}%
\providecommand \@@href[1]{\endgroup#1\@@endlink}%
\providecommand \@sanitize@url [0]{\catcode `\\12\catcode `\$12\catcode
  `\&12\catcode `\#12\catcode `\^12\catcode `\_12\catcode `\%12\relax}%
\providecommand \@@startlink[1]{}%
\providecommand \@@endlink[0]{}%
\providecommand \url  [0]{\begingroup\@sanitize@url \@url }%
\providecommand \@url [1]{\endgroup\@href {#1}{\urlprefix }}%
\providecommand \urlprefix  [0]{URL }%
\providecommand \Eprint [0]{\href }%
\providecommand \doibase [0]{https://doi.org/}%
\providecommand \selectlanguage [0]{\@gobble}%
\providecommand \bibinfo  [0]{\@secondoftwo}%
\providecommand \bibfield  [0]{\@secondoftwo}%
\providecommand \translation [1]{[#1]}%
\providecommand \BibitemOpen [0]{}%
\providecommand \bibitemStop [0]{}%
\providecommand \bibitemNoStop [0]{.\EOS\space}%
\providecommand \EOS [0]{\spacefactor3000\relax}%
\providecommand \BibitemShut  [1]{\csname bibitem#1\endcsname}%
\let\auto@bib@innerbib\@empty
\bibitem [{\citenamefont {Keimer}\ \emph {et~al.}(2015)\citenamefont {Keimer},
  \citenamefont {Kivelson}, \citenamefont {Norman}, \citenamefont {Uchida},\
  and\ \citenamefont {Zaanen}}]{Keimer2015}%
  \BibitemOpen
  \bibfield  {author} {\bibinfo {author} {\bibfnamefont {B.}~\bibnamefont
  {Keimer}}, \bibinfo {author} {\bibfnamefont {S.~A.}\ \bibnamefont
  {Kivelson}}, \bibinfo {author} {\bibfnamefont {M.~R.}\ \bibnamefont
  {Norman}}, \bibinfo {author} {\bibfnamefont {S.}~\bibnamefont {Uchida}},\
  and\ \bibinfo {author} {\bibfnamefont {J.}~\bibnamefont {Zaanen}},\
  }\bibfield  {title} {\bibinfo {title} {{From quantum matter to
  high-temperature superconductivity in copper oxides}},\ }\href
  {https://doi.org/10.1038/nature14165} {\bibfield  {journal} {\bibinfo
  {journal} {Nature}\ }\textbf {\bibinfo {volume} {518}},\ \bibinfo {pages}
  {179} (\bibinfo {year} {2015})}\BibitemShut {NoStop}%
\bibitem [{\citenamefont {Peierls}(1955)}]{Peierls1955}%
  \BibitemOpen
  \bibfield  {author} {\bibinfo {author} {\bibfnamefont {R.}~\bibnamefont
  {Peierls}},\ }\href@noop {} {\emph {\bibinfo {title} {{Quantum Theory of
  Solids}}}}\ (\bibinfo  {publisher} {Oxford University},\ \bibinfo {year}
  {1955})\BibitemShut {NoStop}%
\bibitem [{\citenamefont {Gr\"uner}(1994)}]{Gruner1994}%
  \BibitemOpen
  \bibfield  {author} {\bibinfo {author} {\bibfnamefont {G.}~\bibnamefont
  {Gr\"uner}},\ }\href {https://doi.org/10.1201/9780429501012} {\emph {\bibinfo
  {title} {{Density Waves in Solids}}}}\ (\bibinfo  {publisher} {Boca Raton},\
  \bibinfo {year} {1994})\BibitemShut {NoStop}%
\bibitem [{\citenamefont {Johannes}\ and\ \citenamefont
  {Mazin}(2008)}]{Johannes2008}%
  \BibitemOpen
  \bibfield  {author} {\bibinfo {author} {\bibfnamefont {M.~D.}\ \bibnamefont
  {Johannes}}\ and\ \bibinfo {author} {\bibfnamefont {I.~I.}\ \bibnamefont
  {Mazin}},\ }\bibfield  {title} {\bibinfo {title} {{Fermi surface nesting and
  the origin of charge density waves in metals}},\ }\href
  {https://doi.org/10.1103/PhysRevB.77.165135} {\bibfield  {journal} {\bibinfo
  {journal} {Phys. Rev. B}\ }\textbf {\bibinfo {volume} {77}},\ \bibinfo
  {pages} {165135} (\bibinfo {year} {2008})}\BibitemShut {NoStop}%
\bibitem [{\citenamefont {Zhu}\ \emph {et~al.}(2015)\citenamefont {Zhu},
  \citenamefont {Cao}, \citenamefont {Zhang}, \citenamefont {Plummer},\ and\
  \citenamefont {Guo}}]{Zhu2015}%
  \BibitemOpen
  \bibfield  {author} {\bibinfo {author} {\bibfnamefont {X.}~\bibnamefont
  {Zhu}}, \bibinfo {author} {\bibfnamefont {Y.}~\bibnamefont {Cao}}, \bibinfo
  {author} {\bibfnamefont {J.}~\bibnamefont {Zhang}}, \bibinfo {author}
  {\bibfnamefont {E.~W.}\ \bibnamefont {Plummer}},\ and\ \bibinfo {author}
  {\bibfnamefont {J.}~\bibnamefont {Guo}},\ }\bibfield  {title} {\bibinfo
  {title} {{Classification of charge density waves based on their nature}},\
  }\href {https://doi.org/10.1073/pnas.1424791112} {\bibfield  {journal}
  {\bibinfo  {journal} {Proc. Natl. Acad. Sci.}\ }\textbf {\bibinfo {volume}
  {112}},\ \bibinfo {pages} {2367} (\bibinfo {year} {2015})}\BibitemShut
  {NoStop}%
\bibitem [{\citenamefont {DiMasi}\ \emph {et~al.}(1995)\citenamefont {DiMasi},
  \citenamefont {Aronson}, \citenamefont {Mansfield}, \citenamefont {Foran},\
  and\ \citenamefont {Lee}}]{Dimasi1995}%
  \BibitemOpen
  \bibfield  {author} {\bibinfo {author} {\bibfnamefont {E.}~\bibnamefont
  {DiMasi}}, \bibinfo {author} {\bibfnamefont {M.~C.}\ \bibnamefont {Aronson}},
  \bibinfo {author} {\bibfnamefont {J.~F.}\ \bibnamefont {Mansfield}}, \bibinfo
  {author} {\bibfnamefont {B.}~\bibnamefont {Foran}},\ and\ \bibinfo {author}
  {\bibfnamefont {S.}~\bibnamefont {Lee}},\ }\bibfield  {title} {\bibinfo
  {title} {Chemical pressure and charge-density waves in rare-earth
  tritellurides},\ }\href {https://doi.org/10.1103/PhysRevB.52.14516}
  {\bibfield  {journal} {\bibinfo  {journal} {Phys. Rev. B}\ }\textbf {\bibinfo
  {volume} {52}},\ \bibinfo {pages} {14516} (\bibinfo {year}
  {1995})}\BibitemShut {NoStop}%
\bibitem [{\citenamefont {Malliakas}\ \emph {et~al.}(2005)\citenamefont
  {Malliakas}, \citenamefont {Billinge}, \citenamefont {Kim},\ and\
  \citenamefont {Kanatzidis}}]{Malliakas2005}%
  \BibitemOpen
  \bibfield  {author} {\bibinfo {author} {\bibfnamefont {C.}~\bibnamefont
  {Malliakas}}, \bibinfo {author} {\bibfnamefont {S.~J.~L.}\ \bibnamefont
  {Billinge}}, \bibinfo {author} {\bibfnamefont {H.~J.}\ \bibnamefont {Kim}},\
  and\ \bibinfo {author} {\bibfnamefont {M.~G.}\ \bibnamefont {Kanatzidis}},\
  }\bibfield  {title} {\bibinfo {title} {{Square Nets of Tellurium: Rare-Earth
  Dependent Variation in the Charge-Density Wave of RETe$_3$ (RE = Rare-Earth
  Element)}},\ }\href {https://doi.org/10.1021/ja0505292} {\bibfield  {journal}
  {\bibinfo  {journal} {J. Am. Chem. Soc.}\ }\textbf {\bibinfo {volume}
  {127}},\ \bibinfo {pages} {6510} (\bibinfo {year} {2005})}\BibitemShut
  {NoStop}%
\bibitem [{\citenamefont {Ru}(2008)}]{RuThesis}%
  \BibitemOpen
  \bibfield  {author} {\bibinfo {author} {\bibfnamefont {N.}~\bibnamefont
  {Ru}},\ }\emph {\bibinfo {title} {{Charge Density Wave Formation in
  Rare-earth Tellurides}}},\ \href
  {https://web.stanford.edu/group/fisher/people/Nancy_Ru_thesis.pdf} {\bibinfo
  {type} {{Ph.D. thesis}}},\ \bibinfo  {school} {Stanford University}, \bibinfo
  {address} {Stanford} (\bibinfo {year} {2008})\BibitemShut {NoStop}%
\bibitem [{\citenamefont {Yumigeta}\ \emph {et~al.}(2021)\citenamefont
  {Yumigeta}, \citenamefont {Qin}, \citenamefont {Li}, \citenamefont {Blei},
  \citenamefont {Attarde}, \citenamefont {Kopas},\ and\ \citenamefont
  {Tongay}}]{Yumigeta2021}%
  \BibitemOpen
  \bibfield  {author} {\bibinfo {author} {\bibfnamefont {K.}~\bibnamefont
  {Yumigeta}}, \bibinfo {author} {\bibfnamefont {Y.}~\bibnamefont {Qin}},
  \bibinfo {author} {\bibfnamefont {H.}~\bibnamefont {Li}}, \bibinfo {author}
  {\bibfnamefont {M.}~\bibnamefont {Blei}}, \bibinfo {author} {\bibfnamefont
  {Y.}~\bibnamefont {Attarde}}, \bibinfo {author} {\bibfnamefont
  {C.}~\bibnamefont {Kopas}},\ and\ \bibinfo {author} {\bibfnamefont
  {S.}~\bibnamefont {Tongay}},\ }\bibfield  {title} {\bibinfo {title}
  {{Advances in Rare‐Earth Tritelluride Quantum Materials: Structure,
  Properties, and Synthesis}},\ }\href {https://doi.org/10.1002/advs.202004762}
  {\bibfield  {journal} {\bibinfo  {journal} {Adv. Sci.}\ }\textbf {\bibinfo
  {volume} {8}},\ \bibinfo {pages} {1} (\bibinfo {year} {2021})}\BibitemShut
  {NoStop}%
\bibitem [{\citenamefont {Brouet}\ \emph {et~al.}(2008)\citenamefont {Brouet},
  \citenamefont {Yang}, \citenamefont {Zhou}, \citenamefont {Hussain},
  \citenamefont {Moore}, \citenamefont {He}, \citenamefont {Lu}, \citenamefont
  {Shen}, \citenamefont {Laverock}, \citenamefont {Dugdale}, \citenamefont
  {Ru},\ and\ \citenamefont {Fisher}}]{Brouet2008}%
  \BibitemOpen
  \bibfield  {author} {\bibinfo {author} {\bibfnamefont {V.}~\bibnamefont
  {Brouet}}, \bibinfo {author} {\bibfnamefont {W.~L.}\ \bibnamefont {Yang}},
  \bibinfo {author} {\bibfnamefont {X.~J.}\ \bibnamefont {Zhou}}, \bibinfo
  {author} {\bibfnamefont {Z.}~\bibnamefont {Hussain}}, \bibinfo {author}
  {\bibfnamefont {R.~G.}\ \bibnamefont {Moore}}, \bibinfo {author}
  {\bibfnamefont {R.}~\bibnamefont {He}}, \bibinfo {author} {\bibfnamefont
  {D.~H.}\ \bibnamefont {Lu}}, \bibinfo {author} {\bibfnamefont {Z.~X.}\
  \bibnamefont {Shen}}, \bibinfo {author} {\bibfnamefont {J.}~\bibnamefont
  {Laverock}}, \bibinfo {author} {\bibfnamefont {S.~B.}\ \bibnamefont
  {Dugdale}}, \bibinfo {author} {\bibfnamefont {N.}~\bibnamefont {Ru}},\ and\
  \bibinfo {author} {\bibfnamefont {I.~R.}\ \bibnamefont {Fisher}},\ }\bibfield
   {title} {\bibinfo {title} {{Angle-resolved photoemission study of the
  evolution of band structure and charge density wave properties in
  \emph{\uppercase{R}}{T}e$_3$ (\emph{\uppercase{R}}={Y}, {L}a, {C}e, {S}m,
  {G}d, {T}b, and {D}y)}},\ }\href {https://doi.org/10.1103/PhysRevB.77.235104}
  {\bibfield  {journal} {\bibinfo  {journal} {Phys. Rev. B}\ }\textbf {\bibinfo
  {volume} {77}},\ \bibinfo {pages} {235104} (\bibinfo {year}
  {2008})}\BibitemShut {NoStop}%
\bibitem [{\citenamefont {Gweon}\ \emph {et~al.}(1998)\citenamefont {Gweon},
  \citenamefont {Denlinger}, \citenamefont {Clack}, \citenamefont {Allen},
  \citenamefont {Olson}, \citenamefont {DiMasi}, \citenamefont {Aronson},
  \citenamefont {Foran},\ and\ \citenamefont {Lee}}]{Gweon1998}%
  \BibitemOpen
  \bibfield  {author} {\bibinfo {author} {\bibfnamefont {G.-H.}\ \bibnamefont
  {Gweon}}, \bibinfo {author} {\bibfnamefont {J.~D.}\ \bibnamefont
  {Denlinger}}, \bibinfo {author} {\bibfnamefont {J.~A.}\ \bibnamefont
  {Clack}}, \bibinfo {author} {\bibfnamefont {J.~W.}\ \bibnamefont {Allen}},
  \bibinfo {author} {\bibfnamefont {C.~G.}\ \bibnamefont {Olson}}, \bibinfo
  {author} {\bibfnamefont {E.}~\bibnamefont {DiMasi}}, \bibinfo {author}
  {\bibfnamefont {M.~C.}\ \bibnamefont {Aronson}}, \bibinfo {author}
  {\bibfnamefont {B.}~\bibnamefont {Foran}},\ and\ \bibinfo {author}
  {\bibfnamefont {S.}~\bibnamefont {Lee}},\ }\bibfield  {title} {\bibinfo
  {title} {{Direct observation of complete Fermi surface, imperfect nesting,
  and gap anisotropy in the high-temperature incommensurate charge-density-wave
  compound SmTe$_3$}},\ }\href {https://doi.org/10.1103/PhysRevLett.81.886}
  {\bibfield  {journal} {\bibinfo  {journal} {Phys. Rev. Lett.}\ }\textbf
  {\bibinfo {volume} {81}},\ \bibinfo {pages} {886} (\bibinfo {year}
  {1998})}\BibitemShut {NoStop}%
\bibitem [{\citenamefont {Iyeiri}\ \emph {et~al.}(2003)\citenamefont {Iyeiri},
  \citenamefont {Okumura}, \citenamefont {Michioka},\ and\ \citenamefont
  {Suzuki}}]{Iyeiri2003}%
  \BibitemOpen
  \bibfield  {author} {\bibinfo {author} {\bibfnamefont {Y.}~\bibnamefont
  {Iyeiri}}, \bibinfo {author} {\bibfnamefont {T.}~\bibnamefont {Okumura}},
  \bibinfo {author} {\bibfnamefont {C.}~\bibnamefont {Michioka}},\ and\
  \bibinfo {author} {\bibfnamefont {K.}~\bibnamefont {Suzuki}},\ }\bibfield
  {title} {\bibinfo {title} {{Magnetic properties of rare-earth metal
  tritellurides $R$Te$_\text{3}$ ($R$ = Ce, Pr, Nd, Gd, Dy)}},\ }\href
  {https://doi.org/10.1103/PhysRevB.67.144417} {\bibfield  {journal} {\bibinfo
  {journal} {Phys. Rev. B}\ }\textbf {\bibinfo {volume} {67}},\ \bibinfo
  {pages} {144417} (\bibinfo {year} {2003})}\BibitemShut {NoStop}%
\bibitem [{\citenamefont {Lv}\ \emph {et~al.}(2022)\citenamefont {Lv},
  \citenamefont {Zong}, \citenamefont {Wu}, \citenamefont {Rozhkov},
  \citenamefont {Fine}, \citenamefont {Chen}, \citenamefont {Hashimoto},
  \citenamefont {Lu}, \citenamefont {Li}, \citenamefont {Huang}, \citenamefont
  {Ruff}, \citenamefont {Walko}, \citenamefont {Chen}, \citenamefont {Hwang},
  \citenamefont {Su}, \citenamefont {Shen}, \citenamefont {Wang}, \citenamefont
  {Han}, \citenamefont {Po}, \citenamefont {Wang}, \citenamefont
  {Jarillo-Herrero}, \citenamefont {Wang}, \citenamefont {Zhou}, \citenamefont
  {Sun}, \citenamefont {Wen}, \citenamefont {Shen}, \citenamefont {Wang},\ and\
  \citenamefont {Gedik}}]{LvPRL2021}%
  \BibitemOpen
  \bibfield  {author} {\bibinfo {author} {\bibfnamefont {B.~Q.}\ \bibnamefont
  {Lv}}, \bibinfo {author} {\bibfnamefont {A.}~\bibnamefont {Zong}}, \bibinfo
  {author} {\bibfnamefont {D.}~\bibnamefont {Wu}}, \bibinfo {author}
  {\bibfnamefont {A.~V.}\ \bibnamefont {Rozhkov}}, \bibinfo {author}
  {\bibfnamefont {B.~V.}\ \bibnamefont {Fine}}, \bibinfo {author}
  {\bibfnamefont {S.-D.}\ \bibnamefont {Chen}}, \bibinfo {author}
  {\bibfnamefont {M.}~\bibnamefont {Hashimoto}}, \bibinfo {author}
  {\bibfnamefont {D.-H.}\ \bibnamefont {Lu}}, \bibinfo {author} {\bibfnamefont
  {M.}~\bibnamefont {Li}}, \bibinfo {author} {\bibfnamefont {Y.-B.}\
  \bibnamefont {Huang}}, \bibinfo {author} {\bibfnamefont {J.~P.~C.}\
  \bibnamefont {Ruff}}, \bibinfo {author} {\bibfnamefont {D.~A.}\ \bibnamefont
  {Walko}}, \bibinfo {author} {\bibfnamefont {Z.~H.}\ \bibnamefont {Chen}},
  \bibinfo {author} {\bibfnamefont {I.}~\bibnamefont {Hwang}}, \bibinfo
  {author} {\bibfnamefont {Y.}~\bibnamefont {Su}}, \bibinfo {author}
  {\bibfnamefont {X.}~\bibnamefont {Shen}}, \bibinfo {author} {\bibfnamefont
  {X.}~\bibnamefont {Wang}}, \bibinfo {author} {\bibfnamefont {F.}~\bibnamefont
  {Han}}, \bibinfo {author} {\bibfnamefont {H.~C.}\ \bibnamefont {Po}},
  \bibinfo {author} {\bibfnamefont {Y.}~\bibnamefont {Wang}}, \bibinfo {author}
  {\bibfnamefont {P.}~\bibnamefont {Jarillo-Herrero}}, \bibinfo {author}
  {\bibfnamefont {X.}~\bibnamefont {Wang}}, \bibinfo {author} {\bibfnamefont
  {H.}~\bibnamefont {Zhou}}, \bibinfo {author} {\bibfnamefont {C.-J.}\
  \bibnamefont {Sun}}, \bibinfo {author} {\bibfnamefont {H.}~\bibnamefont
  {Wen}}, \bibinfo {author} {\bibfnamefont {Z.-X.}\ \bibnamefont {Shen}},
  \bibinfo {author} {\bibfnamefont {N.~L.}\ \bibnamefont {Wang}},\ and\
  \bibinfo {author} {\bibfnamefont {N.}~\bibnamefont {Gedik}},\ }\bibfield
  {title} {\bibinfo {title} {Unconventional hysteretic transition in a charge
  density wave},\ }\href {https://doi.org/10.1103/PhysRevLett.128.036401}
  {\bibfield  {journal} {\bibinfo  {journal} {Phys. Rev. Lett.}\ }\textbf
  {\bibinfo {volume} {128}},\ \bibinfo {pages} {036401} (\bibinfo {year}
  {2022})}\BibitemShut {NoStop}%
\bibitem [{\citenamefont {Wu}\ \emph {et~al.}(2019)\citenamefont {Wu},
  \citenamefont {Liu}, \citenamefont {Chen}, \citenamefont {Zhong},
  \citenamefont {Su}, \citenamefont {Shi}, \citenamefont {Tong}, \citenamefont
  {Xu}, \citenamefont {Gao},\ and\ \citenamefont {Wang}}]{Wu2019}%
  \BibitemOpen
  \bibfield  {author} {\bibinfo {author} {\bibfnamefont {D.}~\bibnamefont
  {Wu}}, \bibinfo {author} {\bibfnamefont {Q.~M.}\ \bibnamefont {Liu}},
  \bibinfo {author} {\bibfnamefont {S.~L.}\ \bibnamefont {Chen}}, \bibinfo
  {author} {\bibfnamefont {G.~Y.}\ \bibnamefont {Zhong}}, \bibinfo {author}
  {\bibfnamefont {J.}~\bibnamefont {Su}}, \bibinfo {author} {\bibfnamefont
  {L.~Y.}\ \bibnamefont {Shi}}, \bibinfo {author} {\bibfnamefont
  {L.}~\bibnamefont {Tong}}, \bibinfo {author} {\bibfnamefont {G.}~\bibnamefont
  {Xu}}, \bibinfo {author} {\bibfnamefont {P.}~\bibnamefont {Gao}},\ and\
  \bibinfo {author} {\bibfnamefont {N.~L.}\ \bibnamefont {Wang}},\ }\bibfield
  {title} {\bibinfo {title} {{Layered semiconductor ${\mathrm{EuTe}}_{4}$ with
  charge density wave order in square tellurium sheets}},\ }\href
  {https://doi.org/10.1103/PhysRevMaterials.3.024002} {\bibfield  {journal}
  {\bibinfo  {journal} {Phys. Rev. Mater.}\ }\textbf {\bibinfo {volume} {3}},\
  \bibinfo {pages} {024002} (\bibinfo {year} {2019})}\BibitemShut {NoStop}%
\bibitem [{\citenamefont {Pathak}\ \emph
  {et~al.}(2022{\natexlab{a}})\citenamefont {Pathak}, \citenamefont {Gupta},
  \citenamefont {Mittal},\ and\ \citenamefont {Bansal}}]{Pathak2022}%
  \BibitemOpen
  \bibfield  {author} {\bibinfo {author} {\bibfnamefont {A.}~\bibnamefont
  {Pathak}}, \bibinfo {author} {\bibfnamefont {M.~K.}\ \bibnamefont {Gupta}},
  \bibinfo {author} {\bibfnamefont {R.}~\bibnamefont {Mittal}},\ and\ \bibinfo
  {author} {\bibfnamefont {D.}~\bibnamefont {Bansal}},\ }\bibfield  {title}
  {\bibinfo {title} {Orbital- and atom-dependent linear dispersion across the
  fermi level induces charge density wave instability in ${\text{eute}}_{4}$},\
  }\href {https://doi.org/10.1103/PhysRevB.105.035120} {\bibfield  {journal}
  {\bibinfo  {journal} {Phys. Rev. B}\ }\textbf {\bibinfo {volume} {105}},\
  \bibinfo {pages} {035120} (\bibinfo {year} {2022}{\natexlab{a}})}\BibitemShut
  {NoStop}%
\bibitem [{\citenamefont {Zhang}\ \emph {et~al.}(2023)\citenamefont {Zhang},
  \citenamefont {Shi}, \citenamefont {Zhai}, \citenamefont {Zhao},
  \citenamefont {Du}, \citenamefont {Zhou}, \citenamefont {Gu}, \citenamefont
  {Xu}, \citenamefont {Li}, \citenamefont {Guo}, \citenamefont {Liu},
  \citenamefont {Chen}, \citenamefont {Mo}, \citenamefont {Kim}, \citenamefont
  {Cacho}, \citenamefont {Yu}, \citenamefont {Li}, \citenamefont {Chen},
  \citenamefont {Chu},\ and\ \citenamefont {Yang}}]{ZhangQQ2023}%
  \BibitemOpen
  \bibfield  {author} {\bibinfo {author} {\bibfnamefont {Q.~Q.}\ \bibnamefont
  {Zhang}}, \bibinfo {author} {\bibfnamefont {Y.}~\bibnamefont {Shi}}, \bibinfo
  {author} {\bibfnamefont {K.~Y.}\ \bibnamefont {Zhai}}, \bibinfo {author}
  {\bibfnamefont {W.~X.}\ \bibnamefont {Zhao}}, \bibinfo {author}
  {\bibfnamefont {X.}~\bibnamefont {Du}}, \bibinfo {author} {\bibfnamefont
  {J.~S.}\ \bibnamefont {Zhou}}, \bibinfo {author} {\bibfnamefont
  {X.}~\bibnamefont {Gu}}, \bibinfo {author} {\bibfnamefont {R.~Z.}\
  \bibnamefont {Xu}}, \bibinfo {author} {\bibfnamefont {Y.~D.}\ \bibnamefont
  {Li}}, \bibinfo {author} {\bibfnamefont {Y.~F.}\ \bibnamefont {Guo}},
  \bibinfo {author} {\bibfnamefont {Z.~K.}\ \bibnamefont {Liu}}, \bibinfo
  {author} {\bibfnamefont {C.}~\bibnamefont {Chen}}, \bibinfo {author}
  {\bibfnamefont {S.-K.}\ \bibnamefont {Mo}}, \bibinfo {author} {\bibfnamefont
  {T.~K.}\ \bibnamefont {Kim}}, \bibinfo {author} {\bibfnamefont
  {C.}~\bibnamefont {Cacho}}, \bibinfo {author} {\bibfnamefont {J.~W.}\
  \bibnamefont {Yu}}, \bibinfo {author} {\bibfnamefont {W.}~\bibnamefont {Li}},
  \bibinfo {author} {\bibfnamefont {Y.~L.}\ \bibnamefont {Chen}}, \bibinfo
  {author} {\bibfnamefont {J.-H.}\ \bibnamefont {Chu}},\ and\ \bibinfo {author}
  {\bibfnamefont {L.~X.}\ \bibnamefont {Yang}},\ }\bibfield  {title} {\bibinfo
  {title} {{Thermal hysteretic behavior and negative magnetoresistance in the
  charge density wave material ${\mathrm{EuTe}}_{4}$}},\ }\href
  {https://doi.org/10.1103/PhysRevB.107.115141} {\bibfield  {journal} {\bibinfo
   {journal} {Phys. Rev. B}\ }\textbf {\bibinfo {volume} {107}},\ \bibinfo
  {pages} {115141} (\bibinfo {year} {2023})}\BibitemShut {NoStop}%
\bibitem [{\citenamefont {Zhang}\ \emph {et~al.}(2022)\citenamefont {Zhang},
  \citenamefont {Wu}, \citenamefont {Yuan}, \citenamefont {Zhang},
  \citenamefont {Liu}, \citenamefont {Liu}, \citenamefont {Zhang},
  \citenamefont {Song}, \citenamefont {Zhao}, \citenamefont {Wu}, \citenamefont
  {Liu}, \citenamefont {Chen}, \citenamefont {Ye}, \citenamefont {Cui},
  \citenamefont {Sun}, \citenamefont {Tang}, \citenamefont {He}, \citenamefont
  {Liu}, \citenamefont {Duan}, \citenamefont {Guo},\ and\ \citenamefont
  {Meng}}]{Zhangchen2022}%
  \BibitemOpen
  \bibfield  {author} {\bibinfo {author} {\bibfnamefont {C.}~\bibnamefont
  {Zhang}}, \bibinfo {author} {\bibfnamefont {Q.-Y.}\ \bibnamefont {Wu}},
  \bibinfo {author} {\bibfnamefont {Y.-H.}\ \bibnamefont {Yuan}}, \bibinfo
  {author} {\bibfnamefont {X.}~\bibnamefont {Zhang}}, \bibinfo {author}
  {\bibfnamefont {H.}~\bibnamefont {Liu}}, \bibinfo {author} {\bibfnamefont
  {Z.-T.}\ \bibnamefont {Liu}}, \bibinfo {author} {\bibfnamefont {H.-Y.}\
  \bibnamefont {Zhang}}, \bibinfo {author} {\bibfnamefont {J.-J.}\ \bibnamefont
  {Song}}, \bibinfo {author} {\bibfnamefont {Y.-Z.}\ \bibnamefont {Zhao}},
  \bibinfo {author} {\bibfnamefont {F.-Y.}\ \bibnamefont {Wu}}, \bibinfo
  {author} {\bibfnamefont {S.-Y.}\ \bibnamefont {Liu}}, \bibinfo {author}
  {\bibfnamefont {B.}~\bibnamefont {Chen}}, \bibinfo {author} {\bibfnamefont
  {X.-Q.}\ \bibnamefont {Ye}}, \bibinfo {author} {\bibfnamefont {S.-T.}\
  \bibnamefont {Cui}}, \bibinfo {author} {\bibfnamefont {Z.}~\bibnamefont
  {Sun}}, \bibinfo {author} {\bibfnamefont {X.-F.}\ \bibnamefont {Tang}},
  \bibinfo {author} {\bibfnamefont {J.}~\bibnamefont {He}}, \bibinfo {author}
  {\bibfnamefont {H.-Y.}\ \bibnamefont {Liu}}, \bibinfo {author} {\bibfnamefont
  {Y.-X.}\ \bibnamefont {Duan}}, \bibinfo {author} {\bibfnamefont {Y.-F.}\
  \bibnamefont {Guo}},\ and\ \bibinfo {author} {\bibfnamefont {J.-Q.}\
  \bibnamefont {Meng}},\ }\bibfield  {title} {\bibinfo {title} {{Angle-resolved
  photoemission spectroscopy study of charge density wave order in the layered
  semiconductor ${\text{EuTe}}_{4}$}},\ }\href
  {https://doi.org/10.1103/PhysRevB.106.L201108} {\bibfield  {journal}
  {\bibinfo  {journal} {Phys. Rev. B}\ }\textbf {\bibinfo {volume} {106}},\
  \bibinfo {pages} {L201108} (\bibinfo {year} {2022})}\BibitemShut {NoStop}%
\bibitem [{\citenamefont {Rathore}\ \emph {et~al.}(2023)\citenamefont
  {Rathore}, \citenamefont {Pathak}, \citenamefont {Gupta}, \citenamefont
  {Mittal}, \citenamefont {Kulkarni}, \citenamefont {Thamizhavel},
  \citenamefont {Singhal}, \citenamefont {Said},\ and\ \citenamefont
  {Bansal}}]{Rathore2023}%
  \BibitemOpen
  \bibfield  {author} {\bibinfo {author} {\bibfnamefont {R.}~\bibnamefont
  {Rathore}}, \bibinfo {author} {\bibfnamefont {A.}~\bibnamefont {Pathak}},
  \bibinfo {author} {\bibfnamefont {M.~K.}\ \bibnamefont {Gupta}}, \bibinfo
  {author} {\bibfnamefont {R.}~\bibnamefont {Mittal}}, \bibinfo {author}
  {\bibfnamefont {R.}~\bibnamefont {Kulkarni}}, \bibinfo {author}
  {\bibfnamefont {A.}~\bibnamefont {Thamizhavel}}, \bibinfo {author}
  {\bibfnamefont {H.}~\bibnamefont {Singhal}}, \bibinfo {author} {\bibfnamefont
  {A.~H.}\ \bibnamefont {Said}},\ and\ \bibinfo {author} {\bibfnamefont
  {D.}~\bibnamefont {Bansal}},\ }\bibfield  {title} {\bibinfo {title}
  {{Evolution of static charge density wave order, amplitude mode dynamics, and
  suppression of Kohn anomalies at the hysteretic transition in
  ${\mathrm{EuTe}}_{4}$}},\ }\href
  {https://doi.org/10.1103/PhysRevB.107.024101} {\bibfield  {journal} {\bibinfo
   {journal} {Phys. Rev. B}\ }\textbf {\bibinfo {volume} {107}},\ \bibinfo
  {pages} {024101} (\bibinfo {year} {2023})}\BibitemShut {NoStop}%
\bibitem [{\citenamefont {Liu}\ \emph {et~al.}(2023)\citenamefont {Liu},
  \citenamefont {Wu}, \citenamefont {Wu}, \citenamefont {Han}, \citenamefont
  {Peng}, \citenamefont {Yuan}, \citenamefont {Cheng}, \citenamefont {Li},
  \citenamefont {Hu}, \citenamefont {Yue}, \citenamefont {Xu}, \citenamefont
  {Ding}, \citenamefont {Lu}, \citenamefont {Li}, \citenamefont {Zhang},
  \citenamefont {Lv}, \citenamefont {Zong}, \citenamefont {Su}, \citenamefont
  {Gedik}, \citenamefont {Yin}, \citenamefont {Dong},\ and\ \citenamefont
  {Wang}}]{Liu2023}%
  \BibitemOpen
  \bibfield  {author} {\bibinfo {author} {\bibfnamefont {Q.~M.}\ \bibnamefont
  {Liu}}, \bibinfo {author} {\bibfnamefont {D.}~\bibnamefont {Wu}}, \bibinfo
  {author} {\bibfnamefont {T.~Y.}\ \bibnamefont {Wu}}, \bibinfo {author}
  {\bibfnamefont {S.~S.}\ \bibnamefont {Han}}, \bibinfo {author} {\bibfnamefont
  {Y.~R.}\ \bibnamefont {Peng}}, \bibinfo {author} {\bibfnamefont {Z.~H.}\
  \bibnamefont {Yuan}}, \bibinfo {author} {\bibfnamefont {Y.~H.}\ \bibnamefont
  {Cheng}}, \bibinfo {author} {\bibfnamefont {B.~H.}\ \bibnamefont {Li}},
  \bibinfo {author} {\bibfnamefont {T.~C.}\ \bibnamefont {Hu}}, \bibinfo
  {author} {\bibfnamefont {L.}~\bibnamefont {Yue}}, \bibinfo {author}
  {\bibfnamefont {S.~X.}\ \bibnamefont {Xu}}, \bibinfo {author} {\bibfnamefont
  {R.~X.}\ \bibnamefont {Ding}}, \bibinfo {author} {\bibfnamefont
  {M.}~\bibnamefont {Lu}}, \bibinfo {author} {\bibfnamefont {R.~S.}\
  \bibnamefont {Li}}, \bibinfo {author} {\bibfnamefont {S.~J.}\ \bibnamefont
  {Zhang}}, \bibinfo {author} {\bibfnamefont {B.~Q.}\ \bibnamefont {Lv}},
  \bibinfo {author} {\bibfnamefont {A.}~\bibnamefont {Zong}}, \bibinfo {author}
  {\bibfnamefont {Y.}~\bibnamefont {Su}}, \bibinfo {author} {\bibfnamefont
  {N.}~\bibnamefont {Gedik}}, \bibinfo {author} {\bibfnamefont {Z.~P.}\
  \bibnamefont {Yin}}, \bibinfo {author} {\bibfnamefont {T.}~\bibnamefont
  {Dong}},\ and\ \bibinfo {author} {\bibfnamefont {N.~L.}\ \bibnamefont
  {Wang}},\ }\href@noop {} {\bibinfo {title} {{Room temperature nonvolatile
  optical control of polar order in a charge density wave}}} (\bibinfo {year}
  {2023}),\ \Eprint {https://arxiv.org/abs/2310.10293} {arXiv:2310.10293}
  \BibitemShut {NoStop}%
\bibitem [{\citenamefont {Gedik}\ and\ \citenamefont
  {Vishik}(2017)}]{Gedik2017}%
  \BibitemOpen
  \bibfield  {author} {\bibinfo {author} {\bibfnamefont {N.}~\bibnamefont
  {Gedik}}\ and\ \bibinfo {author} {\bibfnamefont {I.}~\bibnamefont {Vishik}},\
  }\bibfield  {title} {\bibinfo {title} {{Photoemission of quantum
  materials}},\ }\href {https://doi.org/10.1038/nphys4273} {\bibfield
  {journal} {\bibinfo  {journal} {Nat. Phys.}\ }\textbf {\bibinfo {volume}
  {13}},\ \bibinfo {pages} {1029} (\bibinfo {year} {2017})}\BibitemShut
  {NoStop}%
\bibitem [{\citenamefont {Sobota}\ \emph {et~al.}(2021)\citenamefont {Sobota},
  \citenamefont {He},\ and\ \citenamefont {Shen}}]{Sobota2021}%
  \BibitemOpen
  \bibfield  {author} {\bibinfo {author} {\bibfnamefont {J.~A.}\ \bibnamefont
  {Sobota}}, \bibinfo {author} {\bibfnamefont {Y.}~\bibnamefont {He}},\ and\
  \bibinfo {author} {\bibfnamefont {Z.-X.}\ \bibnamefont {Shen}},\ }\bibfield
  {title} {\bibinfo {title} {{Angle-resolved photoemission studies of quantum
  materials}},\ }\href {https://doi.org/10.1103/RevModPhys.93.025006}
  {\bibfield  {journal} {\bibinfo  {journal} {Rev. Mod. Phys.}\ }\textbf
  {\bibinfo {volume} {93}},\ \bibinfo {pages} {025006} (\bibinfo {year}
  {2021})}\BibitemShut {NoStop}%
\bibitem [{\citenamefont {de~la Torre}\ \emph {et~al.}(2021)\citenamefont
  {de~la Torre}, \citenamefont {Kennes}, \citenamefont {Claassen},
  \citenamefont {Gerber}, \citenamefont {McIver},\ and\ \citenamefont
  {Sentef}}]{RMPultrafast2021}%
  \BibitemOpen
  \bibfield  {author} {\bibinfo {author} {\bibfnamefont {A.}~\bibnamefont
  {de~la Torre}}, \bibinfo {author} {\bibfnamefont {D.~M.}\ \bibnamefont
  {Kennes}}, \bibinfo {author} {\bibfnamefont {M.}~\bibnamefont {Claassen}},
  \bibinfo {author} {\bibfnamefont {S.}~\bibnamefont {Gerber}}, \bibinfo
  {author} {\bibfnamefont {J.~W.}\ \bibnamefont {McIver}},\ and\ \bibinfo
  {author} {\bibfnamefont {M.~A.}\ \bibnamefont {Sentef}},\ }\bibfield  {title}
  {\bibinfo {title} {Colloquium: Nonthermal pathways to ultrafast control in
  quantum materials},\ }\href {https://doi.org/10.1103/RevModPhys.93.041002}
  {\bibfield  {journal} {\bibinfo  {journal} {Rev. Mod. Phys.}\ }\textbf
  {\bibinfo {volume} {93}},\ \bibinfo {pages} {041002} (\bibinfo {year}
  {2021})}\BibitemShut {NoStop}%
\bibitem [{\citenamefont {Boschini}\ \emph {et~al.}(2023)\citenamefont
  {Boschini}, \citenamefont {Zonno},\ and\ \citenamefont
  {Damascelli}}]{Boschini2023}%
  \BibitemOpen
  \bibfield  {author} {\bibinfo {author} {\bibfnamefont {F.}~\bibnamefont
  {Boschini}}, \bibinfo {author} {\bibfnamefont {M.}~\bibnamefont {Zonno}},\
  and\ \bibinfo {author} {\bibfnamefont {A.}~\bibnamefont {Damascelli}},\
  }\href@noop {} {\bibinfo {title} {Time- and angle-resolved photoemission
  studies of quantum materials}} (\bibinfo {year} {2023}),\ \Eprint
  {https://arxiv.org/abs/2309.03935} {arXiv:2309.03935} \BibitemShut {NoStop}%
\bibitem [{Rem()}]{Remark1}%
  \BibitemOpen
  \href@noop {} {}\bibinfo {note} {See Supplemental Material for more
  details.}\BibitemShut {Stop}%
\bibitem [{\citenamefont {Schmitt}\ \emph {et~al.}(2011)\citenamefont
  {Schmitt}, \citenamefont {Kirchmann}, \citenamefont {Bovensiepen},
  \citenamefont {Moore}, \citenamefont {Chu}, \citenamefont {Lu}, \citenamefont
  {Rettig}, \citenamefont {Wolf}, \citenamefont {Fisher},\ and\ \citenamefont
  {Shen}}]{Schmitt_2011}%
  \BibitemOpen
  \bibfield  {author} {\bibinfo {author} {\bibfnamefont {F.}~\bibnamefont
  {Schmitt}}, \bibinfo {author} {\bibfnamefont {P.~S.}\ \bibnamefont
  {Kirchmann}}, \bibinfo {author} {\bibfnamefont {U.}~\bibnamefont
  {Bovensiepen}}, \bibinfo {author} {\bibfnamefont {R.~G.}\ \bibnamefont
  {Moore}}, \bibinfo {author} {\bibfnamefont {J.-H.}\ \bibnamefont {Chu}},
  \bibinfo {author} {\bibfnamefont {D.~H.}\ \bibnamefont {Lu}}, \bibinfo
  {author} {\bibfnamefont {L.}~\bibnamefont {Rettig}}, \bibinfo {author}
  {\bibfnamefont {M.}~\bibnamefont {Wolf}}, \bibinfo {author} {\bibfnamefont
  {I.~R.}\ \bibnamefont {Fisher}},\ and\ \bibinfo {author} {\bibfnamefont
  {Z.-X.}\ \bibnamefont {Shen}},\ }\bibfield  {title} {\bibinfo {title}
  {{Ultrafast electron dynamics in the charge density wave material
  TbTe$_3$}},\ }\href {https://doi.org/10.1088/1367-2630/13/6/063022}
  {\bibfield  {journal} {\bibinfo  {journal} {New J. Phys.}\ }\textbf {\bibinfo
  {volume} {13}},\ \bibinfo {pages} {063022} (\bibinfo {year}
  {2011})}\BibitemShut {NoStop}%
\bibitem [{\citenamefont {Kanasaki}\ \emph {et~al.}(2014)\citenamefont
  {Kanasaki}, \citenamefont {Tanimura},\ and\ \citenamefont
  {Tanimura}}]{PRLKanasaki2014}%
  \BibitemOpen
  \bibfield  {author} {\bibinfo {author} {\bibfnamefont {J.}~\bibnamefont
  {Kanasaki}}, \bibinfo {author} {\bibfnamefont {H.}~\bibnamefont {Tanimura}},\
  and\ \bibinfo {author} {\bibfnamefont {K.}~\bibnamefont {Tanimura}},\
  }\bibfield  {title} {\bibinfo {title} {{Imaging energy-, momentum-, and
  time-resolved distributions of photoinjected hot electrons in GaAs}},\ }\href
  {https://doi.org/10.1103/PhysRevLett.113.237401} {\bibfield  {journal}
  {\bibinfo  {journal} {Phys. Rev. Lett.}\ }\textbf {\bibinfo {volume} {113}},\
  \bibinfo {pages} {237401} (\bibinfo {year} {2014})}\BibitemShut {NoStop}%
\bibitem [{\citenamefont {Pathak}\ \emph
  {et~al.}(2022{\natexlab{b}})\citenamefont {Pathak}, \citenamefont {Gupta},
  \citenamefont {Mittal},\ and\ \citenamefont {Bansal}}]{PRB2022}%
  \BibitemOpen
  \bibfield  {author} {\bibinfo {author} {\bibfnamefont {A.}~\bibnamefont
  {Pathak}}, \bibinfo {author} {\bibfnamefont {M.~K.}\ \bibnamefont {Gupta}},
  \bibinfo {author} {\bibfnamefont {R.}~\bibnamefont {Mittal}},\ and\ \bibinfo
  {author} {\bibfnamefont {D.}~\bibnamefont {Bansal}},\ }\bibfield  {title}
  {\bibinfo {title} {{Orbital- and atom-dependent linear dispersion across the
  Fermi level induces charge density wave instability in EuTe$_4$}},\ }\href
  {https://doi.org/10.1103/PhysRevB.105.035120} {\bibfield  {journal} {\bibinfo
   {journal} {Phys. Rev. B}\ }\textbf {\bibinfo {volume} {105}},\ \bibinfo
  {pages} {035120} (\bibinfo {year} {2022}{\natexlab{b}})}\BibitemShut
  {NoStop}%
\bibitem [{\citenamefont {Wen}\ \emph {et~al.}(2021)\citenamefont {Wen},
  \citenamefont {Gao}, \citenamefont {Xie}, \citenamefont {Zhang},
  \citenamefont {Kong}, \citenamefont {Wang}, \citenamefont {Jiang},
  \citenamefont {Luo}, \citenamefont {Li}, \citenamefont {Lu}, \citenamefont
  {Sun},\ and\ \citenamefont {Yan}}]{wenchenh2021}%
  \BibitemOpen
  \bibfield  {author} {\bibinfo {author} {\bibfnamefont {C.}~\bibnamefont
  {Wen}}, \bibinfo {author} {\bibfnamefont {J.}~\bibnamefont {Gao}}, \bibinfo
  {author} {\bibfnamefont {Y.}~\bibnamefont {Xie}}, \bibinfo {author}
  {\bibfnamefont {Q.}~\bibnamefont {Zhang}}, \bibinfo {author} {\bibfnamefont
  {P.}~\bibnamefont {Kong}}, \bibinfo {author} {\bibfnamefont {J.}~\bibnamefont
  {Wang}}, \bibinfo {author} {\bibfnamefont {Y.}~\bibnamefont {Jiang}},
  \bibinfo {author} {\bibfnamefont {X.}~\bibnamefont {Luo}}, \bibinfo {author}
  {\bibfnamefont {J.}~\bibnamefont {Li}}, \bibinfo {author} {\bibfnamefont
  {W.}~\bibnamefont {Lu}}, \bibinfo {author} {\bibfnamefont {Y.-P.}\
  \bibnamefont {Sun}},\ and\ \bibinfo {author} {\bibfnamefont {S.}~\bibnamefont
  {Yan}},\ }\bibfield  {title} {\bibinfo {title} {{Roles of the narrow
  electronic band near the Fermi level in 1$T$-TaS$_\text{2}$-related layered
  materials}},\ }\href {https://doi.org/10.1103/PhysRevLett.126.256402}
  {\bibfield  {journal} {\bibinfo  {journal} {Phys. Rev. Lett.}\ }\textbf
  {\bibinfo {volume} {126}},\ \bibinfo {pages} {256402} (\bibinfo {year}
  {2021})}\BibitemShut {NoStop}%
\bibitem [{\citenamefont {Butler}\ \emph {et~al.}(2020)\citenamefont {Butler},
  \citenamefont {Yoshida}, \citenamefont {Hanaguri},\ and\ \citenamefont
  {Iwasa}}]{Butler2020}%
  \BibitemOpen
  \bibfield  {author} {\bibinfo {author} {\bibfnamefont {C.~J.}\ \bibnamefont
  {Butler}}, \bibinfo {author} {\bibfnamefont {M.}~\bibnamefont {Yoshida}},
  \bibinfo {author} {\bibfnamefont {T.}~\bibnamefont {Hanaguri}},\ and\
  \bibinfo {author} {\bibfnamefont {Y.}~\bibnamefont {Iwasa}},\ }\bibfield
  {title} {\bibinfo {title} {{Mottness versus unit-cell doubling as the driver
  of the insulating state in 1$T$-TaS$_2$}},\ }\href
  {https://doi.org/10.1038/s41467-020-16132-9} {\bibfield  {journal} {\bibinfo
  {journal} {Nat. Commun.}\ }\textbf {\bibinfo {volume} {11}},\ \bibinfo
  {pages} {2477} (\bibinfo {year} {2020})}\BibitemShut {NoStop}%
\bibitem [{\citenamefont {Wang}\ \emph {et~al.}(2020)\citenamefont {Wang},
  \citenamefont {Yao}, \citenamefont {Xin}, \citenamefont {Han}, \citenamefont
  {Wang}, \citenamefont {Chen}, \citenamefont {Cai}, \citenamefont {Li},\ and\
  \citenamefont {Zhang}}]{Wang2020}%
  \BibitemOpen
  \bibfield  {author} {\bibinfo {author} {\bibfnamefont {Y.~D.}\ \bibnamefont
  {Wang}}, \bibinfo {author} {\bibfnamefont {W.~L.}\ \bibnamefont {Yao}},
  \bibinfo {author} {\bibfnamefont {Z.~M.}\ \bibnamefont {Xin}}, \bibinfo
  {author} {\bibfnamefont {T.~T.}\ \bibnamefont {Han}}, \bibinfo {author}
  {\bibfnamefont {Z.~G.}\ \bibnamefont {Wang}}, \bibinfo {author}
  {\bibfnamefont {L.}~\bibnamefont {Chen}}, \bibinfo {author} {\bibfnamefont
  {C.}~\bibnamefont {Cai}}, \bibinfo {author} {\bibfnamefont {Y.}~\bibnamefont
  {Li}},\ and\ \bibinfo {author} {\bibfnamefont {Y.}~\bibnamefont {Zhang}},\
  }\bibfield  {title} {\bibinfo {title} {{Band insulator to Mott insulator
  transition in 1$T$-TaS$_\text{2}$}},\ }\href
  {https://doi.org/10.1038/s41467-020-18040-4} {\bibfield  {journal} {\bibinfo
  {journal} {Nat. Commun.}\ }\textbf {\bibinfo {volume} {11}},\ \bibinfo
  {pages} {4215} (\bibinfo {year} {2020})}\BibitemShut {NoStop}%
\bibitem [{\citenamefont {Ritschel}\ \emph {et~al.}(2018)\citenamefont
  {Ritschel}, \citenamefont {Berger},\ and\ \citenamefont
  {Geck}}]{Ritschel2018}%
  \BibitemOpen
  \bibfield  {author} {\bibinfo {author} {\bibfnamefont {T.}~\bibnamefont
  {Ritschel}}, \bibinfo {author} {\bibfnamefont {H.}~\bibnamefont {Berger}},\
  and\ \bibinfo {author} {\bibfnamefont {J.}~\bibnamefont {Geck}},\ }\bibfield
  {title} {\bibinfo {title} {{Stacking-driven gap formation in layered
  1$T$-TaS$_\text{2}$}},\ }\href {https://doi.org/10.1103/PhysRevB.98.195134}
  {\bibfield  {journal} {\bibinfo  {journal} {Phys. Rev. B}\ }\textbf {\bibinfo
  {volume} {98}},\ \bibinfo {pages} {195134} (\bibinfo {year}
  {2018})}\BibitemShut {NoStop}%
\bibitem [{\citenamefont {Lee}\ \emph {et~al.}(2019)\citenamefont {Lee},
  \citenamefont {Goh},\ and\ \citenamefont {Cho}}]{LeeS2019}%
  \BibitemOpen
  \bibfield  {author} {\bibinfo {author} {\bibfnamefont {S.-H.}\ \bibnamefont
  {Lee}}, \bibinfo {author} {\bibfnamefont {J.~S.}\ \bibnamefont {Goh}},\ and\
  \bibinfo {author} {\bibfnamefont {D.}~\bibnamefont {Cho}},\ }\bibfield
  {title} {\bibinfo {title} {{Origin of the insulating phase and first-order
  metal insulator transition in 1$T$-TaS$_\text{2}$}},\ }\href
  {https://doi.org/10.1103/PhysRevLett.122.106404} {\bibfield  {journal}
  {\bibinfo  {journal} {Phys. Rev. Lett.}\ }\textbf {\bibinfo {volume} {122}},\
  \bibinfo {pages} {106404} (\bibinfo {year} {2019})}\BibitemShut {NoStop}%
\bibitem [{\citenamefont {Halperin}\ and\ \citenamefont
  {Rice}(1968)}]{halperin1968possible}%
  \BibitemOpen
  \bibfield  {author} {\bibinfo {author} {\bibfnamefont {B.}~\bibnamefont
  {Halperin}}\ and\ \bibinfo {author} {\bibfnamefont {T.}~\bibnamefont
  {Rice}},\ }\bibfield  {title} {\bibinfo {title} {Possible anomalies at a
  semimetal-semiconductor transistion},\ }\href@noop {} {\bibfield  {journal}
  {\bibinfo  {journal} {Reviews of Modern Physics}\ }\textbf {\bibinfo {volume}
  {40}},\ \bibinfo {pages} {755} (\bibinfo {year} {1968})}\BibitemShut
  {NoStop}%
\bibitem [{\citenamefont {Damascelli}\ \emph {et~al.}(2003)\citenamefont
  {Damascelli}, \citenamefont {Hussain},\ and\ \citenamefont
  {Shen}}]{Damascelli2003}%
  \BibitemOpen
  \bibfield  {author} {\bibinfo {author} {\bibfnamefont {A.}~\bibnamefont
  {Damascelli}}, \bibinfo {author} {\bibfnamefont {Z.}~\bibnamefont
  {Hussain}},\ and\ \bibinfo {author} {\bibfnamefont {Z.-X.}\ \bibnamefont
  {Shen}},\ }\bibfield  {title} {\bibinfo {title} {{Angle-resolved
  photoemission studies of the cuprate superconductors}},\ }\href
  {https://doi.org/10.1103/RevModPhys.75.473} {\bibfield  {journal} {\bibinfo
  {journal} {Rev. Mod. Phys.}\ }\textbf {\bibinfo {volume} {75}},\ \bibinfo
  {pages} {473} (\bibinfo {year} {2003})}\BibitemShut {NoStop}%
\end{thebibliography}

%

\end{document}